\documentstyle[aps,prb,twocolumn,psfig]{revtex}

\draft
\begin{document}
\twocolumn[\hsize\textwidth\columnwidth\hsize\csname@twocolumnfalse%
\endcsname

\title{Island phases and charge order in two-dimensional manganites}

\author{H. Aliaga,$^1$ B. Normand,$^2$ K. Hallberg,$^1$ M. Avignon,$^3$ 
and B. Alascio$^1$ }

\address{$^1$Instituto Balseiro and Centro Atomico Bariloche, Comision 
Nacional de Energia Atomica, 8400 San Carlos de Bariloche, Argentina.}

\address{$^2$Theoretische Physik III, Elektronische Korrelationen und 
Magnetismus, Institut f\"ur Physik, Universit\"at Augsburg, D-86135 Augsburg, 
Germany.}

\address{$^3$Laboratoire d'Etudes des Propri\'et\'es Electroniques des 
Solides (LEPES), Centre National de la R\'echerche Scientifique, \\ BP 166, 
F-38042 Grenoble Cedex, France.}

\date{\today}

\maketitle

\begin{abstract}

The ferromagnetic Kondo lattice model with an antiferromagnetic 
interaction between localized spins is a minimal description of 
the competing kinetic ($t$) and magnetic ($K$) energy terms which 
generate the rich physics of manganite systems. Motivated by the 
discovery in one dimension of homogeneous ``island phases'', we 
consider the possibility of analogous phases in higher dimensions.
We characterize the phases present at commensurate fillings, and 
consider in detail the effects of phase separation in all filling 
and parameter regimes. We deduce that island and flux phases are 
stable for intermediate values of $K/t$ at the commensurate fillings 
$n$ = 1/4, 1/3, 3/8, and 1/2. We discuss the connection of these 
results to the charge and magnetic ordering observed in a wide 
variety of manganite compounds. 

\end{abstract}

]

\section{Introduction}

Transition-metal manganite compounds have long been known to 
display a broad spectrum of physical properties as a function of 
temperature, filling and counterion composition. While the most 
remarkable of these is the colossal magnetoresistance \cite{rwk} 
observed in the ferromagnetic (FM) phase, the phase diagrams of both 
cubic perovskite and layered manganite materials exhibit a rich 
variety of metallic, insulating, magnetically ordered and, 
apparently, inhomogeneous or phase-separated regions. 

The ferromagnetic Kondo lattice model (FKLM) has been used extensively 
as a minimal model to reproduce the physics responsible for this 
situation. We will study a version of the model which includes a 
Heisenberg interaction between the localized spins. In essence, 
this encapsulates the competition between the 
ferromagnetic polarizing effect of the double-exchange hopping 
term\cite{rzah} ($t$) for mobile carriers in the $e_g$ orbitals of 
Mn$^{3+}$, and the antiferromagnetic (AF) interaction ($K$) between 
the localized spins composed of electrons in the $t_{2g}$ orbitals. 
Treatments of the model with both classical local spins, and with 
fully quantum, $S$ = 1/2 local spins, both return some of the features 
observed among the selection of manganite phase diagrams. A large 
number of authors has worked on many forms of the FKLM, and we will 
present in the following sections only a small selection of references 
relevant to the current approach. 

Following the discovery\cite{rghbaa} in one-dimensional simulations of 
novel ``island phases'' near commensurate values of electron filling in 
the FKLM with strong Hund coupling between localized and conduction 
electrons, we wish here to consider the possibility of higher-dimensional 
generalizations of these phases. By an island phase is meant a spin 
configuration composed of small, regularly arranged, FM islands (clusters 
of 2-4 sites in Ref.~\onlinecite{rghbaa}), with AF local spin orientations 
between islands (Fig.~1). These phases are homogeneous, and near the 
commensurate fillings maximize kinetic energy within each island at 
minimal cost to the magnetic energy, which is favored at the island 
boundaries. Focussing primarily on the problem in two dimensions (2d), 
we wish to establish the possibility that such islands, which may be 
small in one or both directions, remain the most stable phase for certain 
fillings and parameter ratios $K/t$. 

\begin{figure}[hp]
\centerline{\psfig{figure=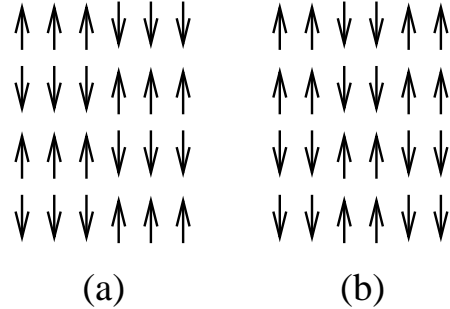,height=4.0cm,angle=0}}
\medskip
\caption{Schematic representations of the island phases $(\pi/3,\pi)$ (a) 
and $(\pi/2,\pi/2)$ (b).}
\end{figure}

A particular motivation for our study is the recent observation of 
charge-ordering phenomena, and more general inhomogeneous charge and 
spin configurations, in a variety of manganite systems. These 
appear in both layered and cubic materials, and at both commensurate 
and incommensurate values of the electron filling set by the counterion 
doping. Some of the earliest 
observations of charge ordering\cite{rcc} were made in 
La$_{1-x}$Sr$_x$MnO$_3$, and were followed by measurements suggesting 
polarons,\cite{ryhnkik} phase separation,\cite{rrdiarmbgda} and paired 
stripe features.\cite{rmcc} Charge order coupled to a structural phase 
transition has been observed in Ba$_{1-x}$Ca$_x$MnO$_3$ at incommensurate 
values of the filling $x$.\cite{rbacc,rlcc} Among hole-doped manganites,  
charge ordering arose at incommensurate filling in Nd$_{1-x}$Sr$_x$MnO$_3$, 
and in a stripe-like configuration at half-filling in 
Pr$_{0.5}$Sr$_{0.5}$MnO$_3$.\cite{rtakmt,rkkytkt} For the latter system, 
the stripe features could be made to ``melt'' in an applied magnetic 
field.\cite{rtakmt} Of most interest in the current context, ordering 
phenomena have also appeared in 2d or layered manganite systems. In   
Sr$_{2-x}$La$_x$MnO$_3$ at low doping, Bao {\it et al.}\cite{rbccc} 
reported charge order, phase separation, and triplet bipolarons. For the 
same system at $x$ = 0.5, Moritomo {\it et al.}\cite{rmnmyoo} related 
charge ordering to lattice effects by substitution for La, and Murakami 
{\it et al.}\cite{rmkktamt} made direct measurements of charge and 
orbital order for the commensurate La member. Finally, we mention also 
the observation\cite{rlmkt} of charge order in the layered 327 compound 
LaSr$_2$Mn$_2$O$_7$. 

The manuscript is organized as follows. In Sec. II we present the 
model in the form we wish to consider, and outline the methods by which 
it is analyzed. In Sec. III we discuss the available means to characterize 
the phases which appear, and illustrate these with examples. Sec. IV 
contains a detailed discussion of the issue of phase separation, and a 
global phase diagram for the augmented FKLM which delimits the regimes of 
interest for island phases. We return in Sec. V to the robust flux and 
island phases, discuss their properties and their charge order, 
and consider their relevance to the above experiments. Sec. VI gives 
a summary and conclusions. 

\section{Model and methods}

We consider the FKLM in the form 
\begin{eqnarray}
H & = & - \sum_{\langle ij \rangle \sigma} t_{ij} \left( c_{i\sigma}^{\dag} 
c_{i\sigma} + {\rm H.c.} \right) - J_H \sum_i {\bf s}_i {\bf .S}_i 
\label{esh} \nonumber \\ & & + K \sum_{\langle ij \rangle} {\bf S}_i 
{\bf .S}_j . 
\end{eqnarray}
Here $c_{i\sigma}^{\dag}$ is the operator creating an electron of spin 
$\sigma$ in the sole $e_g$ orbital; ${\bf s}_i = \sum_{\alpha \beta} 
c_{i\alpha}^{\dag} \sigma_{\alpha \beta} c_{i\beta}$ gives the spin of 
this ``conduction'' electron, and its mobility depends on the orientation 
of the localized $t_{2g}$ spins according to the double-exchange 
mechanism.\cite{rzah}
The second term is the Hund coupling, $J_H > 0$, which favors a FM 
orientation of spins on the same site. Following Refs.~\onlinecite{rmhd,rrhd}, 
we will be concerned with the limit of large $J_H$; while in real systems 
$J_H$ is of the same order as the bandwidth, this simplifying approximation 
has been found to give reasonable results. The limit corresponds to a 
situation where the conduction electron is bound to 
follow the spin texture of the localized system, while anti-aligned 
electrons occupy a band with energy higher by $J_H$. The projecting 
effect of the large Hund coupling allows one to neglect direct Coulomb 
interactions of the $e_g$ electrons. The final term, with $K > 0$, 
expresses the AF interactions between the local $t_{2g}$ spins, whose 
competition with the FM spin alignment required to maximize $e_g$ 
electron kinetic energy (\ref{edet}) generates the intrinsic physics 
of interest in the context of manganite materials. The situation is 
represented schematically in Fig.~2. 

\begin{figure}[hp]
\centerline{\psfig{figure=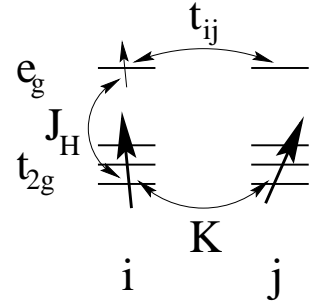,height=4.0cm,angle=0}}
\medskip
\caption{Schematic representation of the Hamiltonian (1) for two sites.}
\end{figure}

We will analyze the model primarily by a classical Monte Carlo (MC) 
procedure for the localized spins, in conjunction with exact 
diagonalization of the conduction electron system.\cite{rdymmhcpf,rym} 
The localized spins are thus taken to be classical, an approximation 
to the true situation of $S = 3/2$ which is found not to invalidate the 
connection to real systems. The conduction electrons are taken to 
occupy a single $e_g$ orbital, or band, and from the condition on $J_H$ 
only one spin projection need be considered. This part of the process 
is the solution of the single-electron problem with hopping set 
consistently by the localized spin configuration. In the limits of 
large $S$ and $J_H$, this is\cite{rmhd}
\begin{equation}
t_{ij} = t \left( \cos \frac{\theta_i}{2} \cos \frac{\theta_j}{2} + 
e^{-(\phi_i - \phi_j)} \sin \frac{\theta_i}{2} \sin \frac{\theta_j}{2} 
\right), 
\label{edet}
\end{equation}
where $\theta_i$ and $\phi_i$ are the polar angles of spin ${\bf S}_i$. 
The resulting energy levels are then filled by the available number of 
electrons in the canonical ensemble. 

The MC simulation proceeds from the FKLM partition function with classical 
spins, 
\begin{equation}
Z = \prod_{i=1}^{N \times N} \int_0^{\pi} d \theta_i \sin \theta_i 
\int_0^{2\pi} d \phi_i \; {\rm Tr} [ \exp ( -\beta H ) ] ,
\label{emcpf}
\end{equation}
where $N$ is the system dimension. Positivity of the integrand 
assures that the sign problem is absent. Updates of the spin 
configuration $\{\theta_i,\phi_i\}$ are accepted or rejected according 
to the Glauber algorithm. Because there are no cases where non-coplanar
spin configurations appear, and because of the large degeneracy of 
coplanar phases, the simulations could be accelerated by fixing 
$\theta_i = \pi/2$, and varying only the angles $\{\phi_i\}$. The number 
of MC steps per site for $N = 8$ is taken as 2000 to equilibrium and 3000
for measurement, while for $N = 12$ the corresponding numbers are 500 
and 1000. Systems of size up to 12$\times$12 are accessible by this 
method, and thus we supplement the MC results by a variety of classical 
analytical considerations, which afford considerable insight and allow 
a detailed assessment of finite-size effects. The simulations may be 
pursued down to temperatures of $T = 0.005t$, which unless otherwise 
stated will be the relevant value for the MC results displayed. This 
temperature is sufficiently 
low that comparison with the zero-temperature, analytical calculations 
is meaningful, and in most cases quantitatively so. The method is the  
same as that used by Dagotto, Yunoki and coworkers in a series 
of papers.\cite{rdymmhcpf,rym,rymd,rmyd,rrdym,ray} We will reproduce some 
of the same results, and comment on the similarities and differences in the 
context of our island phase analyses in what follows. 

\section{Phase characterization}

In this section we will present some results for typical phases which 
emerge from  MC simulations performed at the commensurate fillings 
$n = 1/2$, 1/3, and 1/4, and for the full range of values of $K/t$. 
The results of the simulations for the localized spin system may be 
characterized by three separate but related quantities: the spin 
structure factor 
\begin{equation}
S({\bf k}) = \sum_{i,j} {\bf S}_i {\bf .S}_j e^{i {\bf k.}( {\bf r}_i 
- {\bf r}_j )}, 
\label{essf}
\end{equation}
a histogram of the distribution of angles between all nearest-neighbor 
spin pairs, which we choose to present as a function of $\cos \Theta_{ij}$, 
and a simple ``snapshot'' of the spin configurations at a representative 
step late in the MC process. Note for the histogram that $\Theta_{ij}$ is 
the full angle between spins given by $\cos \Theta_{ij} = ({\bf S}_i 
{\bf .S}_j) / S^2$ for the classical case, and is not to be confused with 
the on-site azimuthal angle $\theta_i$ in Eq.~(\ref{edet}). Finally, one 
may compute in addition the charge distribution function
\begin{equation}
n({\bf k}) = \sum_i n_i e^{i {\bf k \cdot r}_i}, 
\label{ecdf}
\end{equation}
and (by analogy with Eq.~(\ref{essf})) the charge-charge correlation 
function $N({\bf k})$, which we will use in Sec. V when considering 
charge order. 

As a guide to understand the variety of possibilities which is contained 
in these quantities, we first calculate the classical, ground-state energies 
of a multiplicity of possible spin configurations. This may be carried out 
for an infinite 2d system by straightforward extension from the arguments 
presented for the 1d case in Ref.~\onlinecite{rghbaa}. For each spin 
configuration, the magnetic energy per spin is a simple function of the 
average of the angles across each bond, which varies from $2K$ for the 
FM case to $-2K$ for the AF. The kinetic energy at this level is a readily 
calculable function of the spin configuration which varies from 0 in the AF 
case, where all kinetic processes are excluded, to the average energy of 
the 2d nearest-neighbor band $\epsilon_k = - 2 t \left( \cos k_x + \cos 
k_y \right)$, for the relevant band filling, in the FM case where it is 
maximally negative. The results of this exercise are illustrated in Fig.~3 
for $n = 1/2$, in Fig.~4 for $n = 1/3$, and in Fig.~5 for $n = 1/4$. 

All of the phases denoted by $(k\pi/m,l\pi/m)$ have neighboring spins 
only either parallel or antiparallel, in both directions. The rational 
fractions $k/m,l/m$ may be understood as indicating that the spin 
direction turns over $k$ or $l$ times in $2m$ lattice constants. Fig.~1 
shows two small-$m$ possibilities, the $(\pi/3,\pi)$ (a) and $(\pi/2,
\pi/2)$ (b) phases. As a more complex example, the phase $(3\pi/4,\pi)$, 
which appears over a wide range of $K/t$ at filling $n$ = 1/4 (Figs.~5,8), 
would be composed of chains with repeat unit $\uparrow \uparrow \downarrow 
\uparrow \downarrow \downarrow \uparrow \downarrow$ in the $x$-direction, 
and AF alignment in the $y$-direction. In 
addition to these phases, which include the FM $(0,0)$ and AF $(\pi,\pi)$ 
endpoints, we include also the ``flux phase'',\cite{rykm,ray} which will 
be discussed in more detail below, and a ``double spiral'' (DS) phase, by 
which is meant a single phase where the nearest-neighbor spins rotate 
by the same angle $0 \le \Theta \le \pi$ in both $x$ and $y$ directions. 
In this last case, the optimal angle $\Theta$ is obtained by minimizing a 
function of $K/t$, and the double spiral may be expected to be more 
favorable than any variety of single-spiral phases combined with other 
forms of modulation in the transverse direction. Although we have 
considered many possible phases of the above types, in Figs. 3-5 we 
include for clarity only those which are the ground state for some range 
of $K/t$. 

\begin{figure}[hp]
\centerline{\psfig{figure=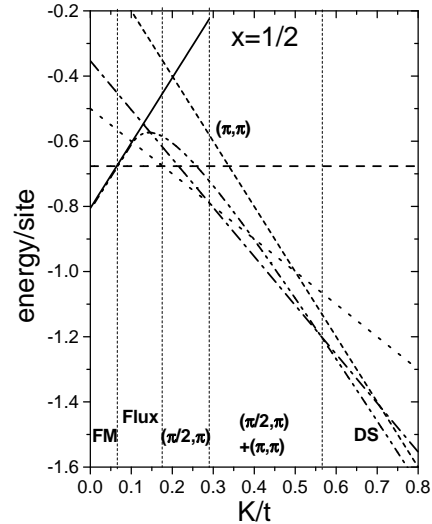,height=7cm,angle=0}}
\medskip
\caption{Energies of selected spin configurations for $n = 1/2$ at all 
values of $K/t$. Note the clear succession of the ground state with 
increasing $K/t$ from flux phase to $(\pi/2,\pi)$ to 
$(\pi/2,\pi)$+$(\pi/,\pi)$. }
\end{figure}

The calculation of all of these phase energies is straightforward. In 
brief, calculation of the only 2d band at $(0,0)$ proceeds as above, 
with the filling determining the chemical potential up to which the 
filled band is integrated. For the 1d structures $(0,l\pi/m)$, one may 
consider the band $\epsilon_k = - 2 t \cos k$ in the continuous 
direction, split appropriately into 2, 3, or 4 (the maximum included 
here) by an equal interchain hopping $t$. Integration over the filled 
parts of these bands up to the chemical potential yields the average 
kinetic energy. For the ``0d'' structures $(k\pi/m,l\pi/m)$, the kinetic 
energy is a simple $m^2/kl$-site diagonalization problem to obtain the 
discrete levels. These phases are particularly favorable when the 
filling exactly matches a large gap in the few-level spectrum, e.g. 
$(\pi/3,\pi)$ for $n = 1/3$ (Fig.~1(a)), or $(\pi/2,\pi/2)$ for $n = 1/4$ 
(Fig.~1(b)). The calculation of the kinetic part for the double-spiral 
phase follows the 2d case above, with reduction of the bandwidth by a 
factor of $\cos \Theta/2$, while the magnetic part varies as $\cos 
\Theta$. We do not find that canted states are favored in these 
considerations. Finally, two special configurations which require 
separate consideration are the $(\pi/2,\pi)$+$(\pi,\pi)$ phase, to 
which we return in Fig.~14, and flux phases. 

\begin{figure}[hp]
\centerline{\psfig{figure=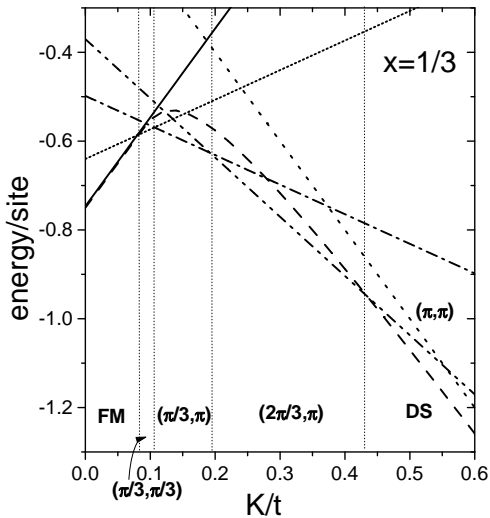,height=7cm,angle=0}}
\medskip
\caption{Energies of selected spin configurations for $n = 1/3$ at all 
values of $K/t$. Note the competition of several phases around $K/t = 
0.1$. }
\end{figure}

Flux phases\cite{rykm} are an important feature of the model in any 
dimension higher than 1. From Eq. (\ref{edet}) it is clear that the 
hopping term also contains a phase factor, and that for certain spin 
textures this phase may differ depending on the path through the lattice 
taken between two points. The simplest flux phase is that appearing at 
half-filling over a broad range of intermediate $K/t$, as discussed in 
Ref.~\onlinecite{ray}, and shown in the snapshot in Fig. 13(c) below. 
The term ``flux phase'' is used here to refer to any spin configuration 
with this non-trivial topological property, which can be quantified by 
a non-zero spin current.\cite{ray} In principle, a variety of flux phases 
may exist, but we have not yet been able to find any others which are 
ground states at any filling. At the analytical level, the semimetallic 
density of states\cite{rykm} of the dispersion
\begin{equation}
\epsilon_{\bf k} = \pm \sqrt{ \cos k_x^2 + \cos k_y^2 } 
\label{efpd}
\end{equation}
of the simplest flux phase,
which is zero precisely at half-filling, accounts for its particularly 
low energy at $n = 1/2$. We will characterize this phase in detail in 
Sec. V.

While these classical, zero-temperature pictures turn out to be rather 
valuable, and also not quantitatively unreasonable, for understanding the 
2d pictures to follow, they are limited by the imagination of the authors 
as further possibilities may not be excluded. We have obtained many of the 
phases proposed in Figs.~3-5 in MC simulations, and the following Figs.~6-8 
illustrate some representative results.

\begin{figure}[hp]
\centerline{\psfig{figure=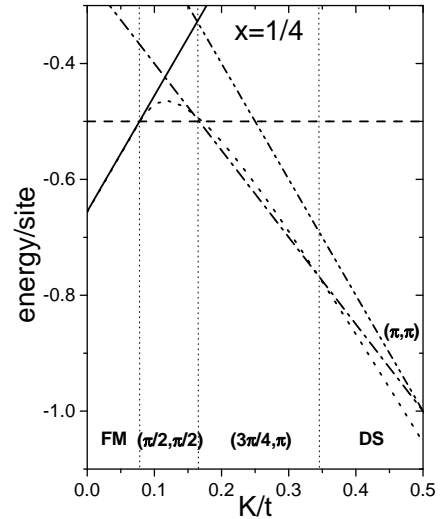,height=7cm,angle=0}}
\medskip
\caption{Energies of selected spin configurations for $n = 1/4$ at all 
values of $K/t$. Note the dominance of the phase $(\pi/2,\pi/2)$ at 
intermediate $K/t$. }
\end{figure}

In Fig.~6 is shown $S({\bf k})$, histogram and snapshot information for 
a phase at filling $n = 1/2$ and for the ratio $K/t = 0.22$. We see a 
single peak in $S({\bf k})$ (Fig.~6(a)) only at $(\pi/2,\pi)$, indicating 
an island phase of FM pairs (the ``islands'') arranged in an AF pattern. 
The histogram (Fig.~6(b)) shows essentially only angles of $0$ and $\pi$, 
ruling out a possible interpretation as a $\pi/2$ spiral in one 
direction; the ratios of angles $0$ to angles $\pi$ is approximately 
1:3 as expected. Finally, the instantaneous spin configuration in Fig.~6(c) 
illustrates that the simulation has in fact converged quite well to the 
expected phase. Comparison with Fig.~3 indicates that for the 2d case, 
the value of $K/t$ for a robust $(\pi/2,\pi)$ phase is that expected 
from the infinite system at $T = 0$. 

Fig.~7 illustrates the same quantities for filling $n = 1/3$ and $K/t = 
0.25$. For this relatively large parameter ratio, the dominant 
$(2\pi/3,\pi)$ phase in $S({\bf k})$ (Fig.~7(a)) consists of AF chains 
with spin configuration $\uparrow \uparrow \downarrow \uparrow \uparrow 
\downarrow \uparrow \uparrow \downarrow \dots$\cite{rghbaa} This is 
one of the primary types of island phase which we will mention again 
in Secs. IV and V. All three parts of Fig.~7 show in addition that this 
phase is not pure in the small-system MC simulation, with spin 
misalignments across the cluster manifest in a broadening of the 
histogram peaks and residual components in $S({\bf k})$. However, 
Fig.~7(b) does demonstrate the absence of intermediate angles which 
might be expected from any kind of spiral phase. 

\begin{figure}[hp]
\medskip
\centerline{\psfig{figure=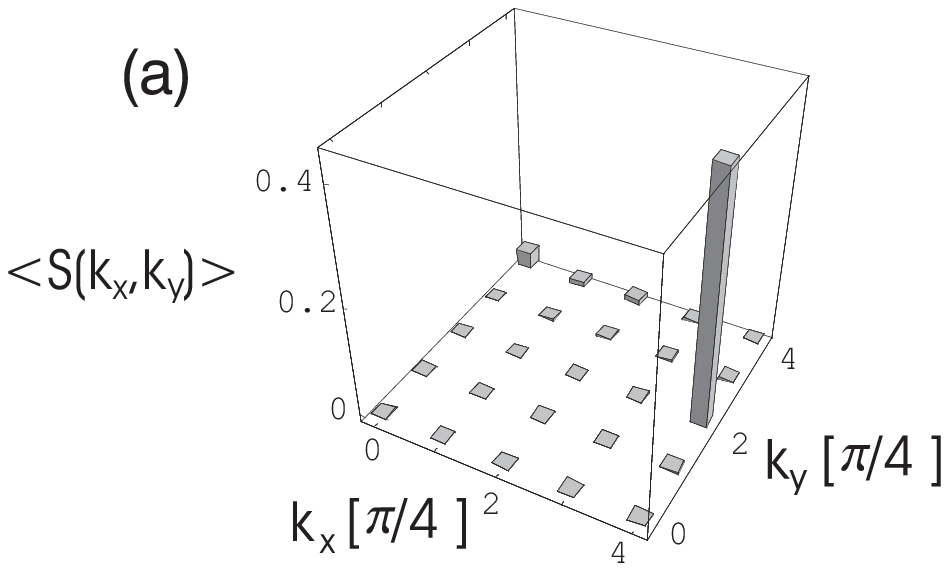,height=4.75cm,angle=0}}
\bigskip
\centerline{\psfig{figure=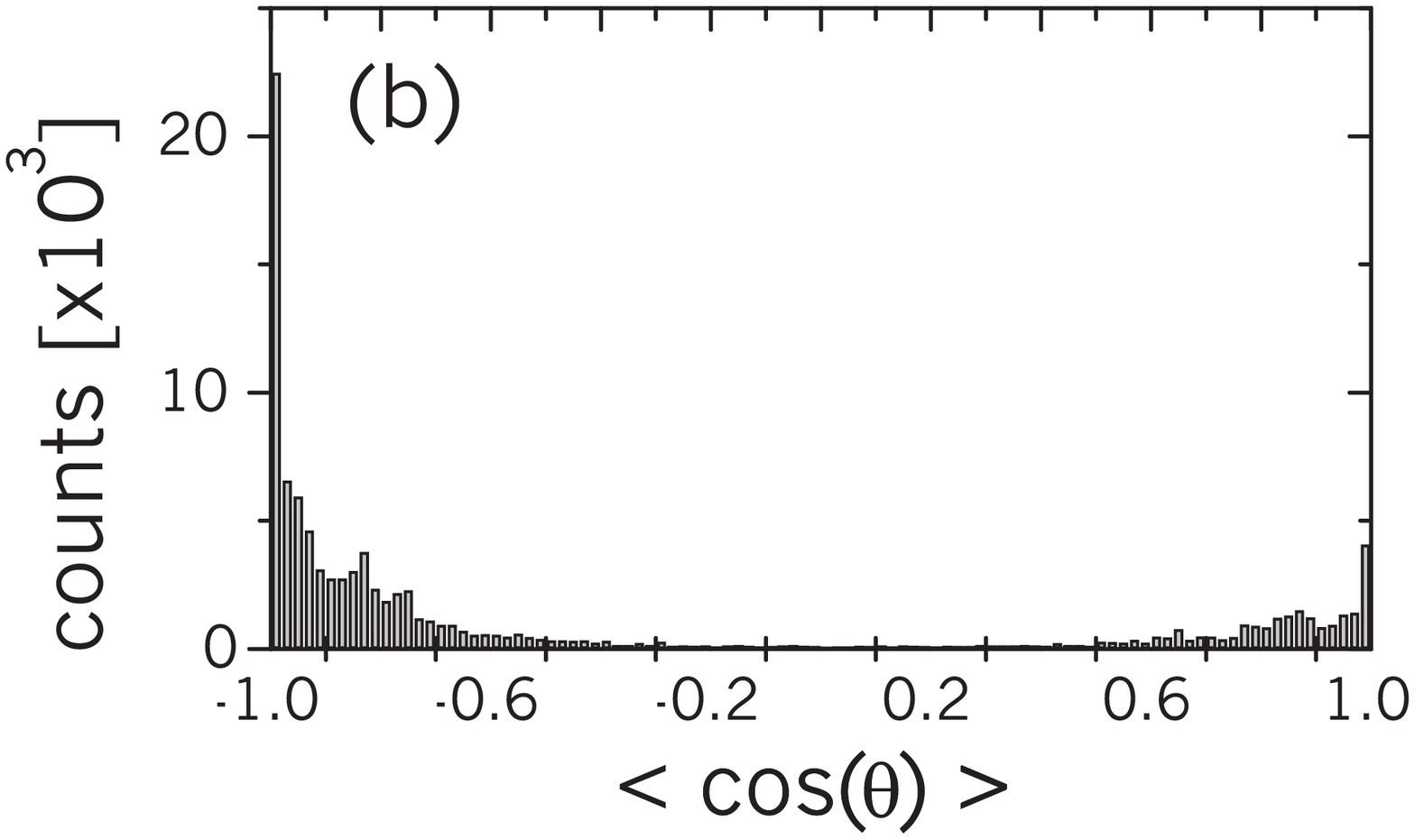,height=4.25cm,angle=0}}
\medskip
\centerline{\psfig{figure=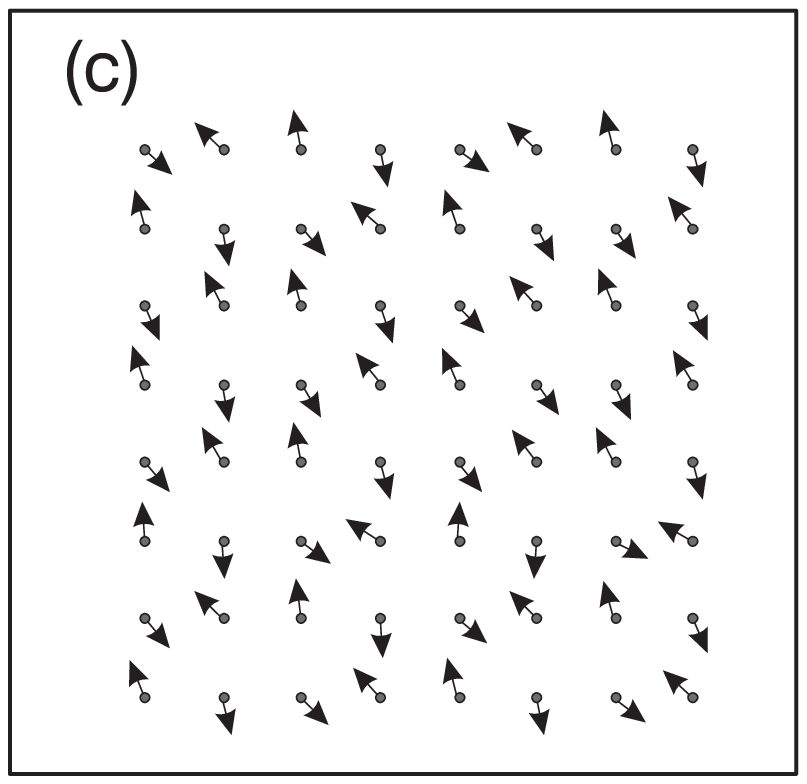,height=5cm,angle=0}}
\medskip
\caption{MC phase for $n = 1/2$ at $K/t$ = 0.22, calculated for an 
8$\times$8 system. (a) Structure factor. (b) Angle histogram. 
(c) Configuration snapshot. }
\end{figure}

Fig.~8 characterizes the phase arising for $n = 1/4$ at $K/t = 0.20$. 
From Fig.~5 we expect the phase $(3\pi/4,\pi)$ as ground state, and 
indeed this is the dominant component in $S({\bf k})$ (Fig.~8(a)). 
The rather stronger admixture of other components arises because the 
chosen value of $K/t$ is close to a phase crossover, and so other 
possible 8$\times$8 phases are not entirely absent. These are not 
reflected in the histogram (Fig.~8(b)) because all the pure phases 
present have angles of only 0 or $\pi$, but the snapshot (Fig.~8(c)) 
does show a small amount of misalignment between the predominantly 
AF-oriented spins. We note that the expected pure configuration 
(see below Fig.~3) remains rather hard to observe in Fig.~8(c), and 
ascribe this to the mixing problem, and to the effects of fluctuations 
on the small-cluster MC calculation. This example illustrates both the 
need for careful consideration of finite-size effects, and the fact that 
for all commensurate fillings there exist regions of $K/t$ (close to the 
line-crossings in Figs.~3-5) where the MC results show strong mixtures 
of different phases. We note in passing that for all fillings we find 
pure FM phases at small but finite $K/t$ ratios, in accord with 
zero-temperature, infinite-system expectations based on Figs.~3-5. 
These straightforward cases are not shown here. At large values of 
$K/t$, small-cluster calculations are unable to access the double 
spiral phase, and show instead the AF. We defer a 
more detailed characterization of the most interesting phases in these 
figures, namely the flux phase at $n = 1/2$, the $(\pi/3,\pi)$ phase at 
$n = 1/3$, and the $(\pi/2,\pi/2)$ phase at $n = 1/4$, until Sec. V, 
after addressing the question of phase separation. 

The results of Fig.~6-8 were obtained for small systems, where finite-size 
effects are of paramount importance. For fillings $n = 1/2$ and $n = 1/4$ 
we may compare 4$\times$4 with 8$\times$8 MC results, and for $n = 1/3$ 
6$\times$6 with 12$\times$12. These comparisons give already a good 
indication of where, for example, certain of the many possible phases 
are anomalously favored by the location of the chemical potential relative 
to a gap between sets of degenerate states. Even more valuable information 
is provided by comparison with the infinite-system results: these may be 
augmented by performing the same calculation, placing spins in a fixed 
configuration and deducing the magnetic and kinetic energies, for the 
system sizes 4$\times$4 to 12$\times$12 of the simulations (and further 
for 16$\times$16). An effective calibration of the MC results is then 
possible, by which is meant a renormalization to account for effects 
arising only from system size, which is particularly important in 
discussing phase transitions (Sec. V). 
 
On these finite systems we are unable to observe phase transitions, 
which are replaced by crossovers occurring in a finite range of $K/t$. 
As we will show in Sec. V, however, a certain amount of care is required 
in interpreting two-peak features in $S({\bf k})$, because some robust, 
single phases arising at particular values of filling and $K/t$ do indeed 
have more than one characteristic wave vector in small systems. Another 
feature requiring particular attention is the possibility of 
large-unit-cell phases, which cannot be accessed in the MC simulations. 
An example already mentioned is the double spiral, which is 
expected from Figs.~3-5 to be the most favorable phase on approaching 
the AF limit, into which this phase in fact passes continuously. However, 
at intermediate to large values of $K/t$ we must also consider a competing, 
large-unit-cell (large-$m$) phase of the type $((m-k)\pi/m,\pi)$, $k \ll 
m$, with only $0$ and $\pi$ angles between the spins, in which the kinetic 
energy gain comes from spins shared between rare FM pairs in an otherwise 
AF structure. These phases are compared in the next section. 

\begin{figure}[hp]
\medskip
\centerline{\psfig{figure=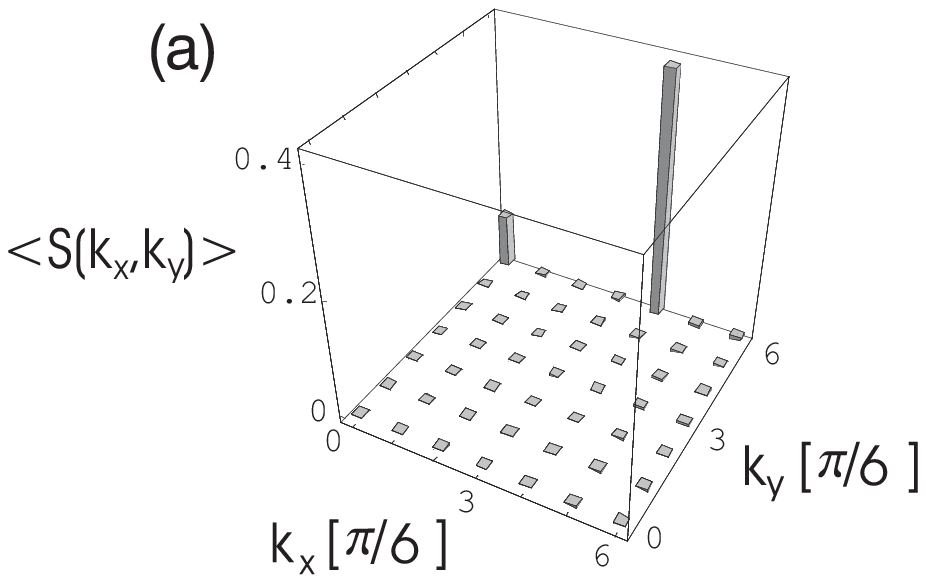,height=4.75cm,angle=0}}
\bigskip
\centerline{\psfig{figure=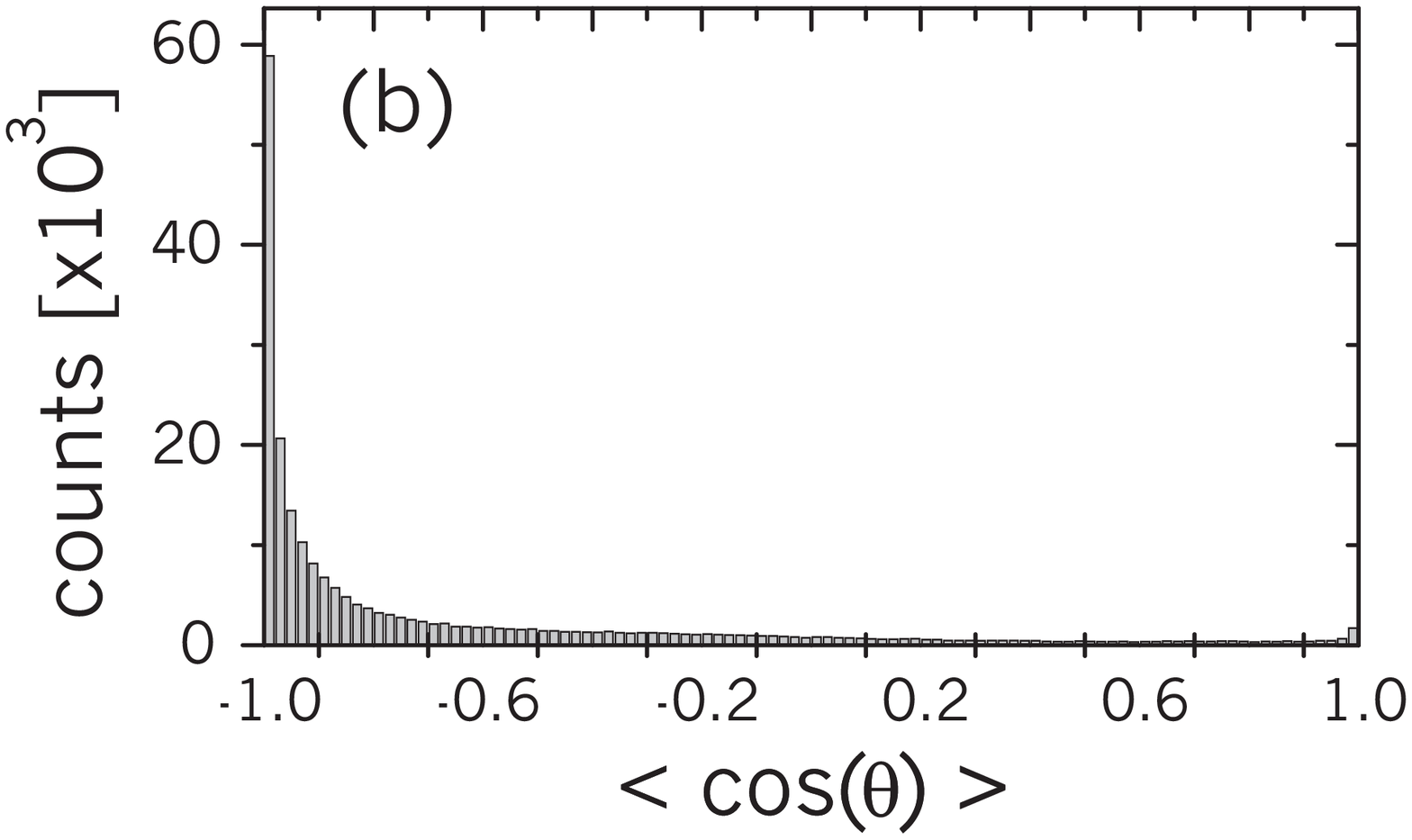,height=4.25cm,angle=0}}
\medskip
\centerline{\psfig{figure=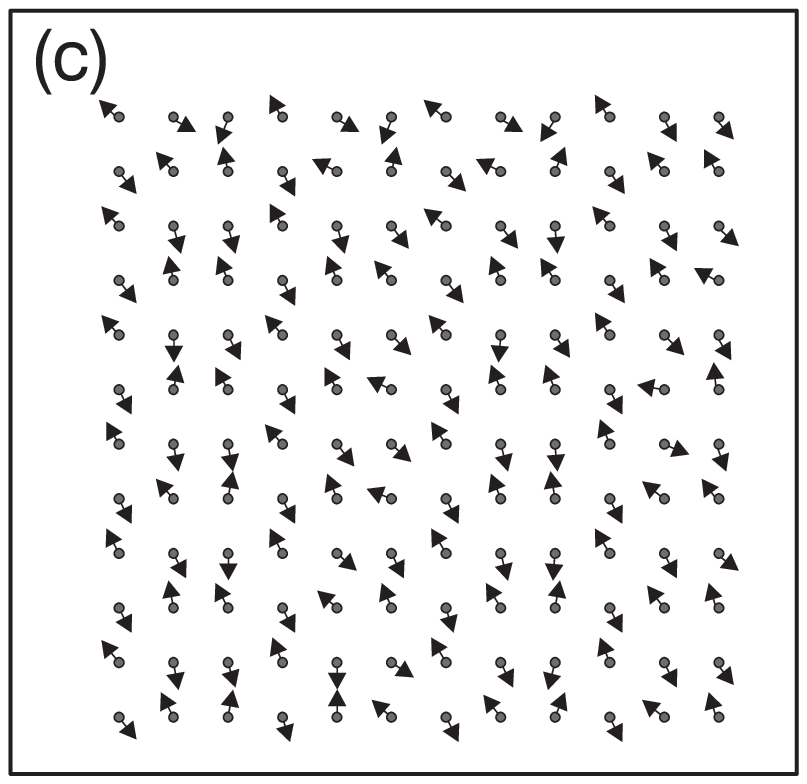,height=5cm,angle=0}}
\medskip
\caption{ MC phase for $n = 1/3$ at $K/t$ = 0.25, calculated for a
12$\times$12 system. (a) Structure factor. (b) Angle histogram. 
(c) Configuration snapshot. }
\end{figure}

To conclude this section, we find that island-like phases are quite 
ubiquitous at all intermediate values of $K/t$ (Figs.~3-5). The FM islands 
may be restricted in one direction, giving rise to stripe-like features, 
or in both to give true islands, depending on the filling. These states 
are also accompanied by flux phases, of non-trivial spin texture, in 
certain parameter regimes. These novel, homogeneous phases arise only 
as a result of the competition between the first and last terms in Eq. 
(\ref{esh}), without recourse to additional physics (a discussion of 
which is deferred to a later section). However, we have worked in a 
canonical ensemble and considered only the energy of the emerging phases 
at zero or the lowest temperatures. We now turn to the question of phase 
separation within the model.

\begin{figure}[hp]
\medskip
\centerline{\psfig{figure=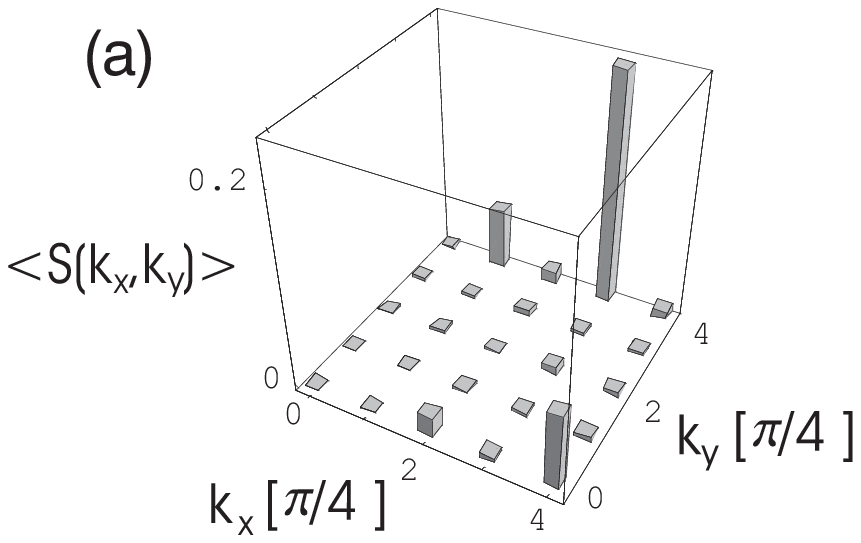,height=4.75cm,angle=0}}
\bigskip
\centerline{\psfig{figure=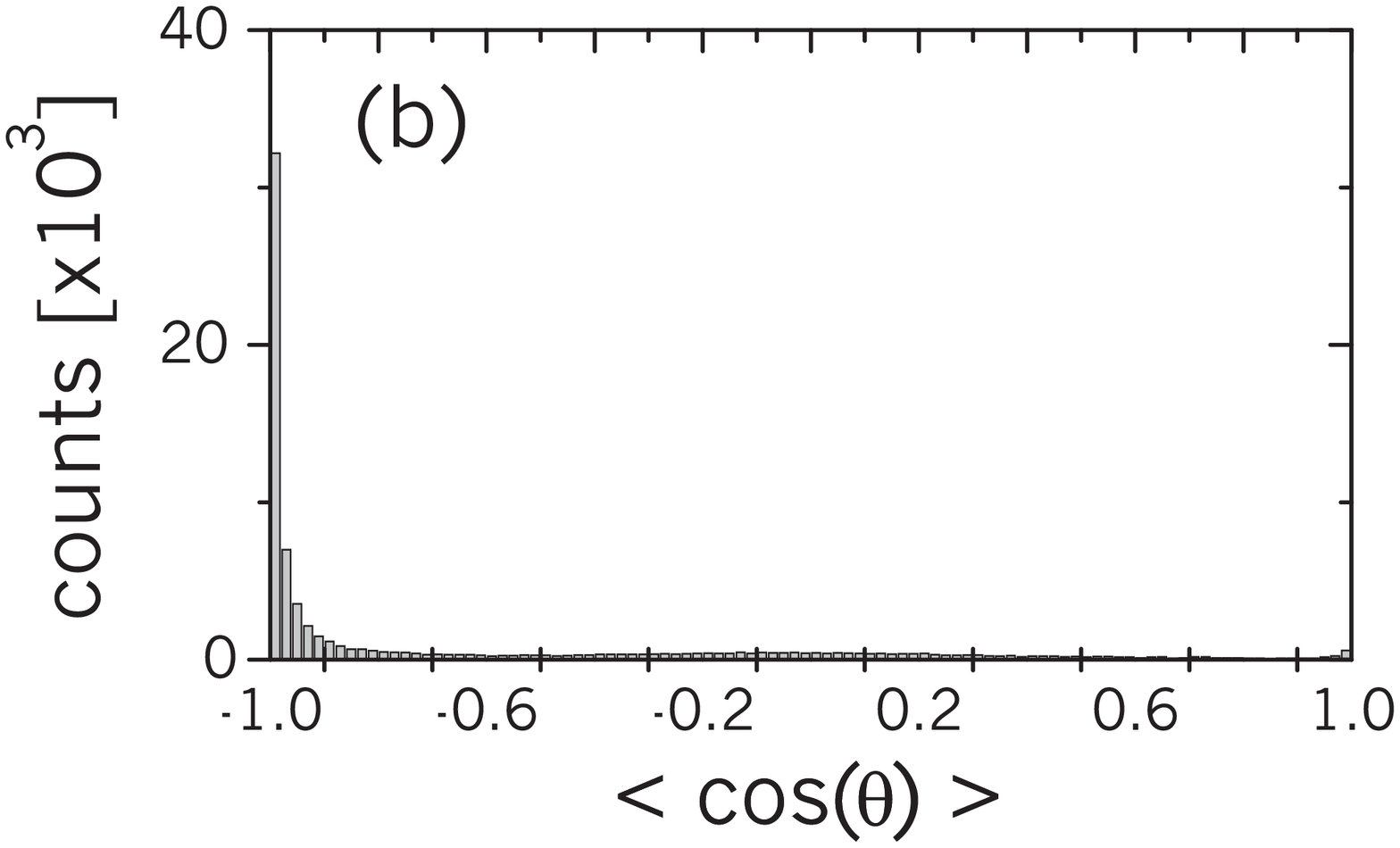,height=4.25cm,angle=0}}
\medskip
\centerline{\psfig{figure=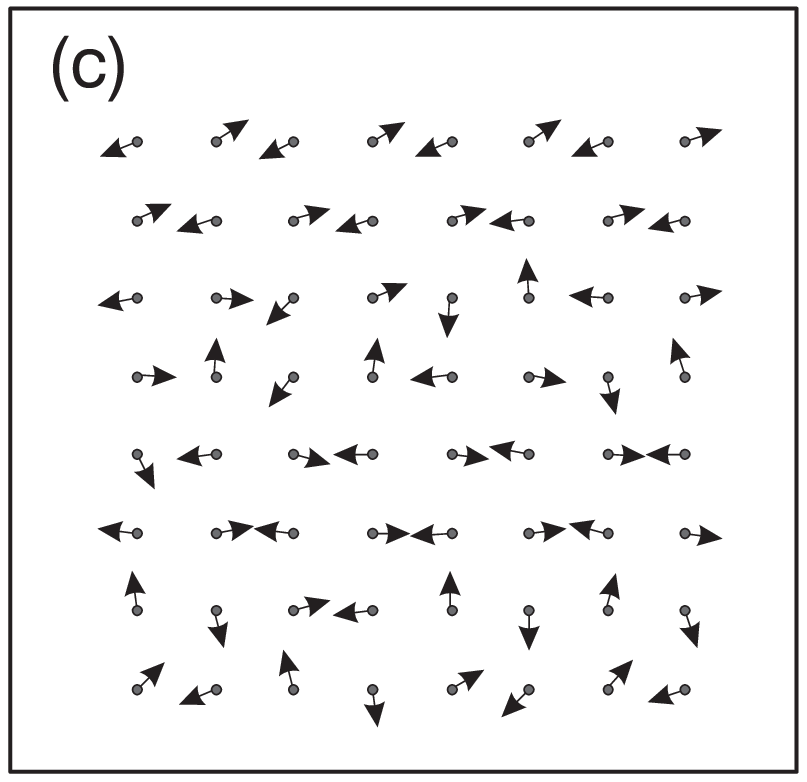,height=5cm,angle=0}}
\medskip
\caption{MC phase for $n = 1/4$ at $K/t$ = 0.20, calculated for an 
8$\times$8 system. (a) Structure factor. (b) Angle histogram. 
(c) Configuration snapshot. }
\end{figure}

\section{Phase separation}

In the previous section we have considered a canonical ensemble, meaning 
fixed particle number, and deduced the ground states on the basis of 
minimal internal energy (or free energy at very low temperature). To 
ensure the global stability of these phases we must consider the possibility 
of their separation into regions of distinct and different filling. This 
propensity has been shown in the same model applied in 1d,\cite{rym} by 
working in a grand canonical ensemble and observing discontinuities in 
filling on varying the chemical potential. Here we choose to characterize 
phase separation from the energy in the canonical ensemble, by observing 
the curvature of this quantity as a function of filling. In Figs.~9-11 are 
shown the energy for fillings $n$ between 0 and 1/2, at low, intermediate 
and higher values of $K/t$. We note that the energy is a symmetrical function 
for $1/2 \le n \le 1$ by electron-hole transformation, and do not comment 
further on this region. In these figures are included data from 
12$\times$12 and 16$\times$16 systems, and infinite-system values 
for the flux and double spiral phases. 

In Fig.~9 we see a convex (up) region at low filling, the implication of 
which is a preference for phase separation into two regions, one of 
zero hole content and the other whose filling $n$ is given by a Maxwell 
construction using the tangent to the concave part of the curve. The 
empty region would have AF spin configuration, while for this low value 
of $K/t$ the partially filled region would be FM. This result confirms 
that phase separation is an important property of the model, and agrees 
qualitatively with Ref.~\onlinecite{rym}. 

\begin{figure}[hp]
\centerline{\psfig{figure=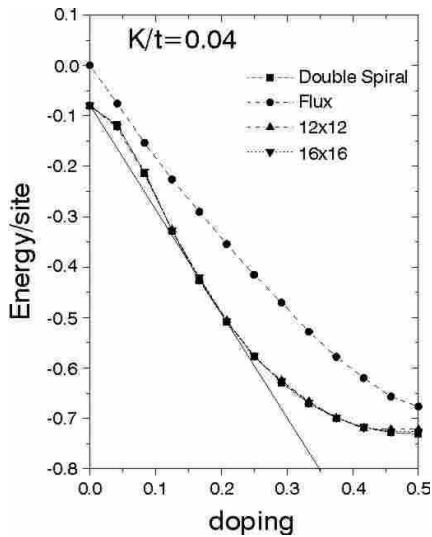,height=7cm,angle=0}}
\medskip
\caption{Energy as a function of filling at fixed $K/t = 0.04$ for a 
variety of phases. The tangent to the curve indicates the regime of 
phase separation by the Maxwell construction. }
\end{figure}

For intermediate $K/t$ (Fig.~10) the situation is more complex. The 
convex regime extends over a much broader range of filling, but the 
``curve'' is much less smooth, as a result of the particularly favorable 
island phases which can be established at the commensurate fillings. In 
fact, Maxwell constructions applied to Fig.~10 yield for this value of 
$K/t$ a separation only into phases $n = 0$ and 1/4, or into $n = 1/4$ 
and $n$ close to 1/2. Inspection of Figs.~3-5 shows that for $K/t = 
0.12$ the phases at these two fillings are particularly robust, whereas 
at $n = 1/3$ a crossing between two phases occurs; by contrast, an 
$n = 1/3$ phase would be expected as an endpoint of such separation 
for $K/t = 0.15$, and indeed emerges (Fig.~12, below). This result 
highlights the dominant r\^ole of the commensurately filled phases, and 
suggests both ``high-contrast'' and ``low-contrast'' phase separation. 
By this is meant in the former case the abrupt split into zero- and 
part-filled regions, and in the latter a finer phase separation for 
certain $K/t$ where incommensurate fillings $1/4 < n < 1/2$ may undergo 
separation into regions with closely neighboring, more commensurate 
fillings. These statements are made systematic in the summary phase diagram 
presented as Fig.~12. 

\begin{figure}[hp]
\centerline{\psfig{figure=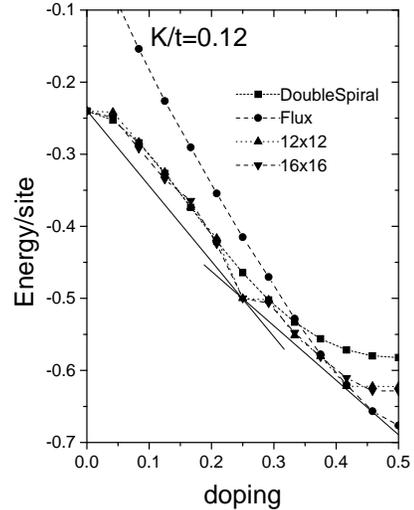,height=7cm,angle=0}}
\medskip
\caption{Energy as a function of filling at fixed $K/t = 0.12$ for 
a variety of phases. The solid lines are Maxwell constructions. }
\end{figure}

At large $K/t$ (Fig. 11) the picture changes again. Here the 
finite-system points for commensurate phases show the intriguing 
feature of lying on a straight line connecting zero- and 1/2-filling. 
These are the $((m-k)\pi/m,\pi)$ phases introduced above, for those 
values of $m$ small enough for the unit cell to fit within the system 
studied. Simple consideration of fixed spin configurations suggests 
that, in principle, phases of arbitrarily large unit-cell size are 
possible, and their energies will fall on the same line. From above, 
the nature of these phases is an AF configuration of spin chains with 
$k$ up-spin and $k$ down-spin pairs contained in an otherwise AF 
system with unit-cell size $2m$. In a fully classical system there 
would be no phase separation with filling in the thermodynamic limit 
at large $K/t$, but instead a continuous evolution of the unit-cell 
dimension to accommodate the added charges. In fact the values of $k$
and $m$ are fixed rather simply by the filling $n$, because the phases 
of this type appearing as the ground state are $((1-n)\pi,\pi)$, and 
their energy is given from the number of FM pairs and AF bonds as 
\begin{equation}
E = - 2 K + n(K - t) 
\label{eluce}
\end{equation}
per site. For the commensurate fillings $n = 1/m = 1/2$, 1/3, and 1/4, 
we recover the island phases of Figs.~3-5. These phases appear to have 
been overlooked in Ref.~\onlinecite{rym}, although the authors were little 
concerned with the high-$K$ regime. 

Our conclusions are summarized in the global phase diagram of Fig.~12. 
The properties of the minimal form of the FKLM (Eq. (\ref{esh})) fall 
broadly into four regions, determined largely by the ratio $K/t$ of the 
super- and double-exchange energy scales. For the lowest values of $K/t$, 
the system separates into AF and FM phases. For small to intermediate 
ratios, $0.08 < K/t < 0.2$, there is large-scale phase separation into 
only the island phases appearing at the commensurate fillings $n$ = 1/4, 
1/3, 3/8, and 1/2. An exception here is the flux phase, which occupies a 
finite doping region close to $n$ = 1/2. We note in passing that within 
our classical formulation, only the FM and flux phases offer the 
possibility of hopping of conduction electrons throughout the system; 
only these phases would have metallic properties, and all others will 
be insulating. 

\begin{figure}[hp]
\centerline{\psfig{figure=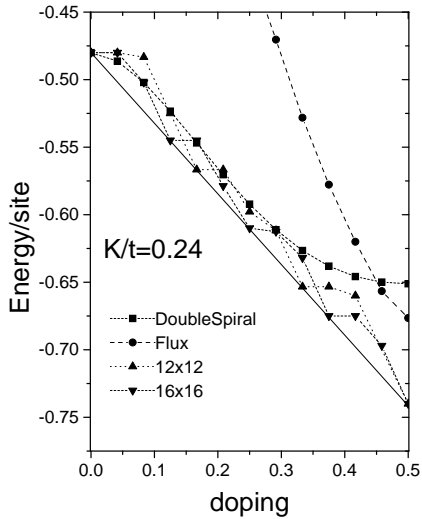,height=7cm,angle=0}}
\medskip
\caption{Energy as a function of filling at fixed $K/t = 0.24$ for a 
variety of phases. The solid line is a guide to the eye. }
\end{figure}

\noindent
For intermediate ratios $0.2 < K/t < 0.28$ we find the 
large-unit-cell phases discussed above. The hierarchy of possible states 
exists across the full doping range only when no competing phase falls 
below the straight-line energy function (Fig.~11, Eq.~(\ref{eluce})) 
for any filling, and it is this condition which sets the limits in 
$K/t$ of the shaded region in Fig.~12. We have marked (vertical dashed 
lines) the small-$m$ phases which are compatible with the finite clusters 
considered, but stress again that from the present calculations we expect 
to find all phases of the form $((1-n) \pi, \pi)$ for the infinite system.
All states in the shaded region are a form of 2-site FM island phase, 
which would show charge-ordering peaks in $N({\bf k})$ (Sec. V), while 
the small-$m$ members at the commensurate fillings provide examples which 
may be studied on small clusters (Figs.~6-8).
At intermediate to large values of $K/t$, the large-unit-cell 
phases are replaced by a wide region of ``high-contrast'' phase separation 
due to the extraordinary stability of the $(\pi/2,\pi)$+$(\pi,\pi)$ phase 
at $n = 1/2$. We have found only this phase, which is considered in more 
detail in the following section, and the AF phase with zero filling, to 
be stable in this regime of $K/t$,but stress that we cannot fully exclude 
the possibility of similar $((m-k)\pi/m,\pi)$+$(\pi,\pi)$ phases at other 
commensurate fillings. A search for these is limited by the available 
cluster size, and remains a topic for future investigation. Finally, at 
large values of $K/t$ we recover the conventional, spiral-ordered DS phase, 
which passes smoothly to an AF phase. 

\begin{figure}[hp]
\medskip
\centerline{\psfig{figure=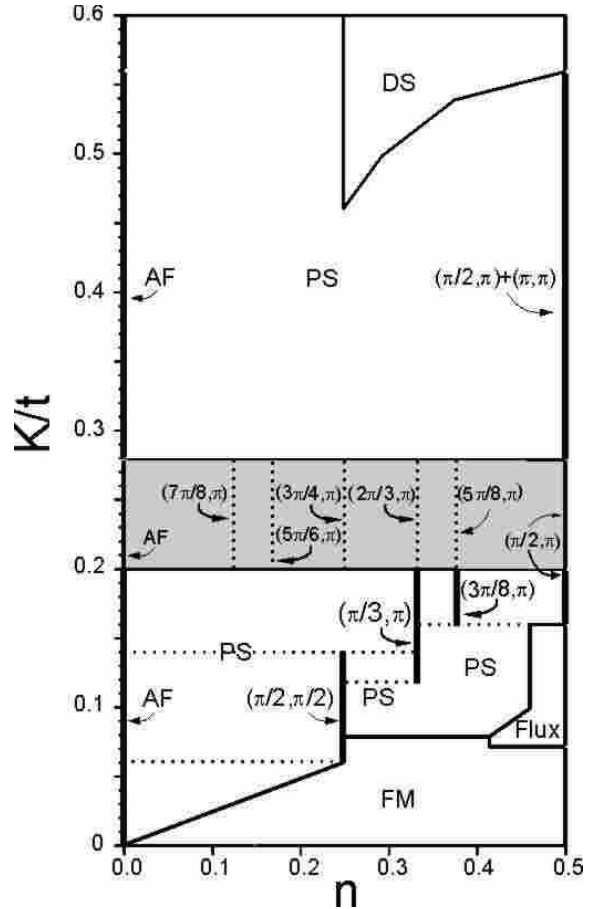,height=12cm,angle=0}}
\medskip
\caption{Phase diagram of augmented FKLM for the full range of filling $n$ 
and ratio $K/t$. PS denotes phase separation, the thick, vertical lines 
the island phases, and the shaded region the regime of large-unit-cell 
phases. }
\end{figure}

\section{Island phases}

With the results of the previous section concerning phase stability and 
separation, we may now turn in more detail to the regime of interest for 
island phases. This is largely limited to the commensurate fillings 
$n = 1/2$, 1/3, and 1/4, and to the parameter range $0.1 < K/t < 0.3$, which 
(Fig.~12) encompasses both the isolated phases which are PS endpoints, and 
the large-unit-cell phases. For $n = 1/2$, this region is dominated first 
by the flux phase, shown in Fig.~13. In Fig 13(a), we see the double-peak 
structure of $S({\bf k})$ with equal weight in $(0,\pi)$ and $(\pi,0)$ 
components which is the hallmark\cite{ray} of this spin configuration. 
Fig.~13(b) provides then a rare example of a phase where the angles between 
neighboring spins are distributed not around the FM and AF configurations, 
but around $\pi/2$; our distribution is narrower than that in 
Ref.~\onlinecite{ray} because of the larger lattice size employed. 
This spin configuration gives rise to a uniform charge distribution with 
no inhomogeneous ordering. Fig.~13(c) shows the real-space spin structure. 

\begin{figure}[hp]
\medskip
\centerline{\psfig{figure=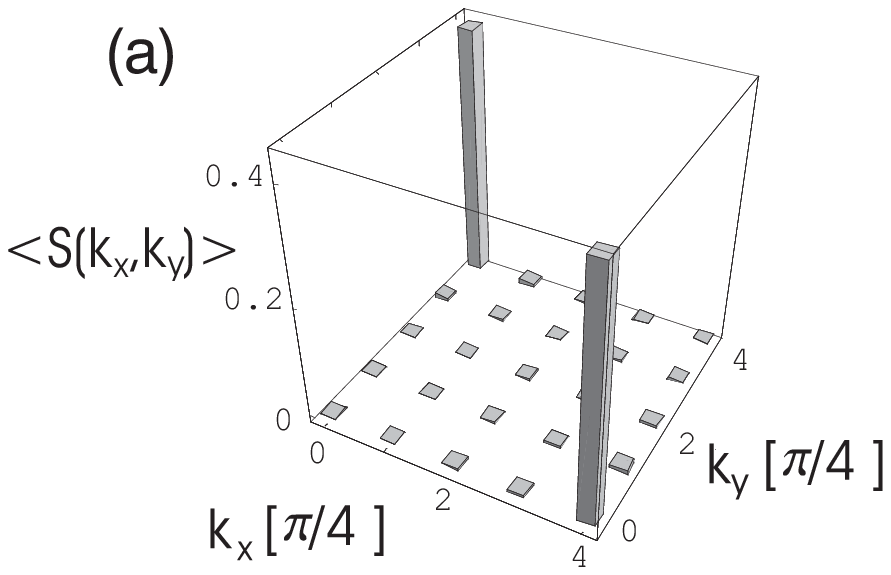,height=4.75cm,angle=0}}
\bigskip
\centerline{\psfig{figure=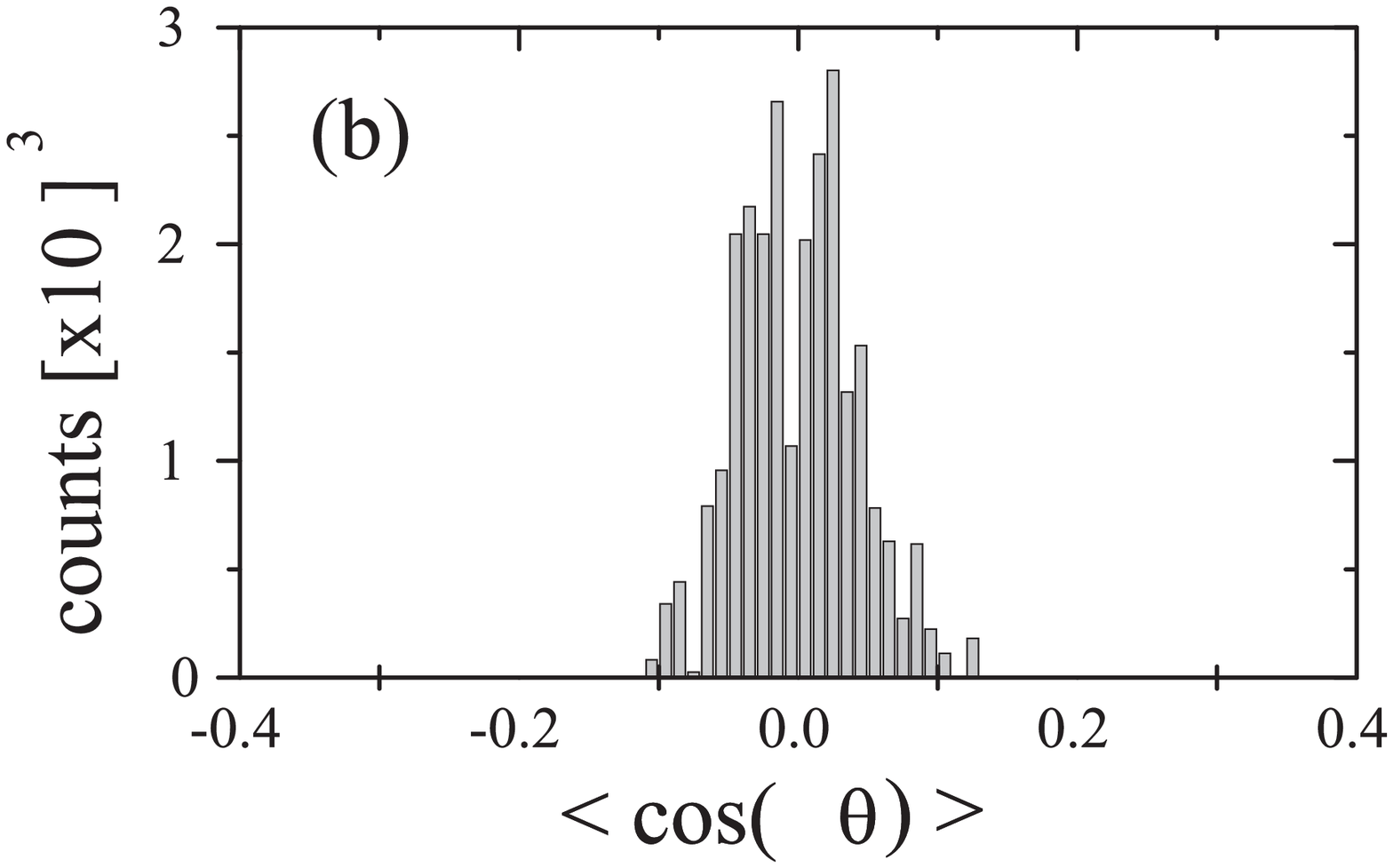,height=4.25cm,angle=0}}
\medskip
\centerline{\psfig{figure=ff12c.eps,height=5cm,angle=0}}
\medskip
\caption{MC results for $n = 1/2$ at $K/t$ = 0.12, characterizing the 
flux phase on an 8$\times$8 lattice. (a) Structure factor. (b) Angle 
histogram. (c) Configuration snapshot. }
\end{figure}

By contrast, for the same filling at larger $K/t$, it is possible to find 
inhomogeneous charge structures. The $(\pi/2,\pi)$ phase of Fig.~6 exists 
as an endpoint both of phase separation and of the large-unit-cell series 
(Fig.~12). In this structure, electrons are delocalized across every 
second bond, or equivalently every FM bond in the $\pi/2$ direction, and 
are much more weakly present on the alternate AF bonds. This simple picture 
implies a stripe-like charge order with wave vector $(\pi,0)$, and the phase 
would give peaks in x-ray diffraction or electron microscopy experiments, 
which measure the charge distribution $n({\bf r})$. However, because 
the charge density $n_i$ is the same on all sites, there is no structure 
in the quantity $n({\bf k})$, which is defined in Eq.~(\ref{ecdf}) and 
readily calculated on a finite cluster. This situation arises only for 
periodicities $\pi/2$ in the spin structure factor; for all higher $m$ 
values, $n({\bf k})$ and $N({\bf k})$ computed from the site charges 
are indeed suitable indicators of charge order. We note briefly here that 
by translational invariance one may in fact expect to find a linear 
superposition of equivalent island phases, with a uniform mean value of 
$n_i$, and a charge order discernible only in $N({\bf k})$. In the 
classical MC simulations we have shown results only for one such phase, 
which is separated by thermal barriers from its degenerate counterparts.

In Fig.~14 we show a 
further stable configuration, which we call the $(\pi/2,\pi)$+$(\pi,\pi)$ 
phase. Here (Fig.~14(a)) there are two peaks in the structure factor, but 
these do not indicate a mixture of phases. While the histogram information 
(Fig.~14(b)) can be used only to rule out intermediate angles, it is the 
instantaneous MC spin configuration (Fig.~14(c)) which reveals the true 
nature of this homogeneous phase. Once again one expects a 1d charge order 
for the same reasons as above. It is this phase, whose energy falls below 
the function given in Eq.~(\ref{eluce}) for $n = 1/2$ and $K/t > 0.28$, 
which breaks the large-unit-cell sequence, and is responsible for the wide 
region of high-contrast PS in the phase diagram of Fig.~12. 

We dwell only briefly on the case of 3/8 filling. The results 
from the previous section show a $(3\pi/8,\pi)$ phase to be a stable 
endpoint in the PS regime, while the large-unit-cell region contains a 
$(5\pi/8,\pi)$ member. The properties of these configurations are 
readily deduced by comparison with the other examples presented, and 
both have charge-ordering wave vectors of $(\pi/4,0)$. Certain 
anomalies have been observed in experiment for filling $n = 3/8$, but 
these appear to be restricted to 3d systems.
Turning to $n = 1/3$, the most robust island phase in the intermediate 
parameter range is $(\pi/3,\pi)$, illustrated schematically in Fig.~1(a), 
and for $K/t$ = 0.15 in Fig.~15. At this value of $K/t$, Fig.~15(a) shows 
a rather strong $(\pi/3,\pi)$ peak, while Fig.~15(b) suggests a 1:2 ratio 
between FM and AF bond angles despite the weak presence of a 
$(\pi/3,2\pi/3)$ component. Fig.~15(c) shows the actual spin structure, 
which gives rise to a charge order at the wave vector $(2\pi/3,0)$, due 
to the higher population of every third site in the $\pi/3$ direction. 
This ordering is present in the site charge distribution function $n_i$, 
which is shown in Fig.~16. From Fig.~16(a) it is clear that the 
charge contrast between the centre and edge sites of each island approaches 
the classical ratio of 2:1.\cite{rghbaa} Very similar results are obtained 
for the ``$((1-n)\pi,\pi)$'' phase $(2\pi/3,\pi)$ as $K/t$ is raised 
beyond 0.2, as already shown in Fig.~7. In this state the charge-ordering 
wave vector remains $(2\pi/3,0)$. We have not been able to find a novel 
flux phase for 1/3 filling which might be a ground state anywhere in the 
intermediate $K/t$ regime.

\begin{figure}[hp]
\medskip
\centerline{\psfig{figure=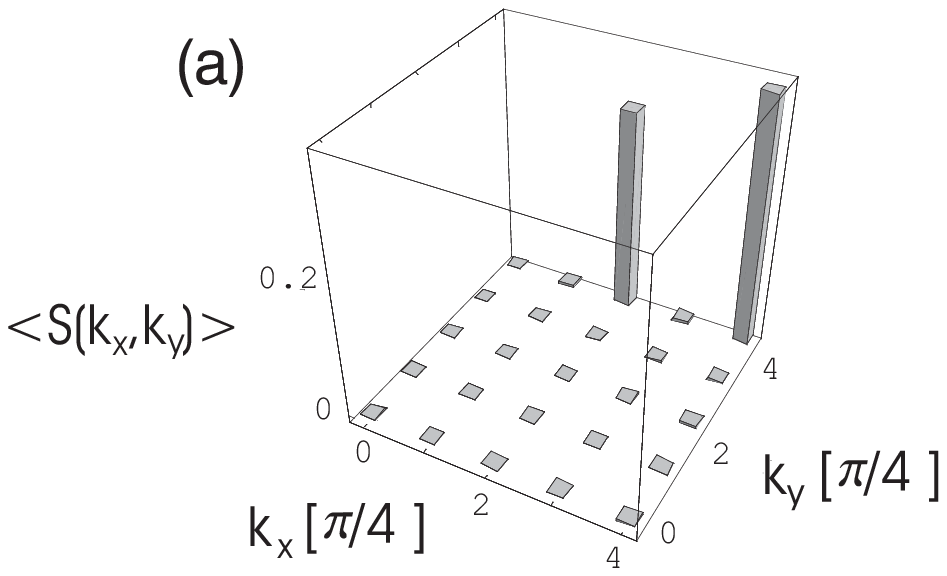,height=4.75cm,angle=0}}
\bigskip
\centerline{\psfig{figure=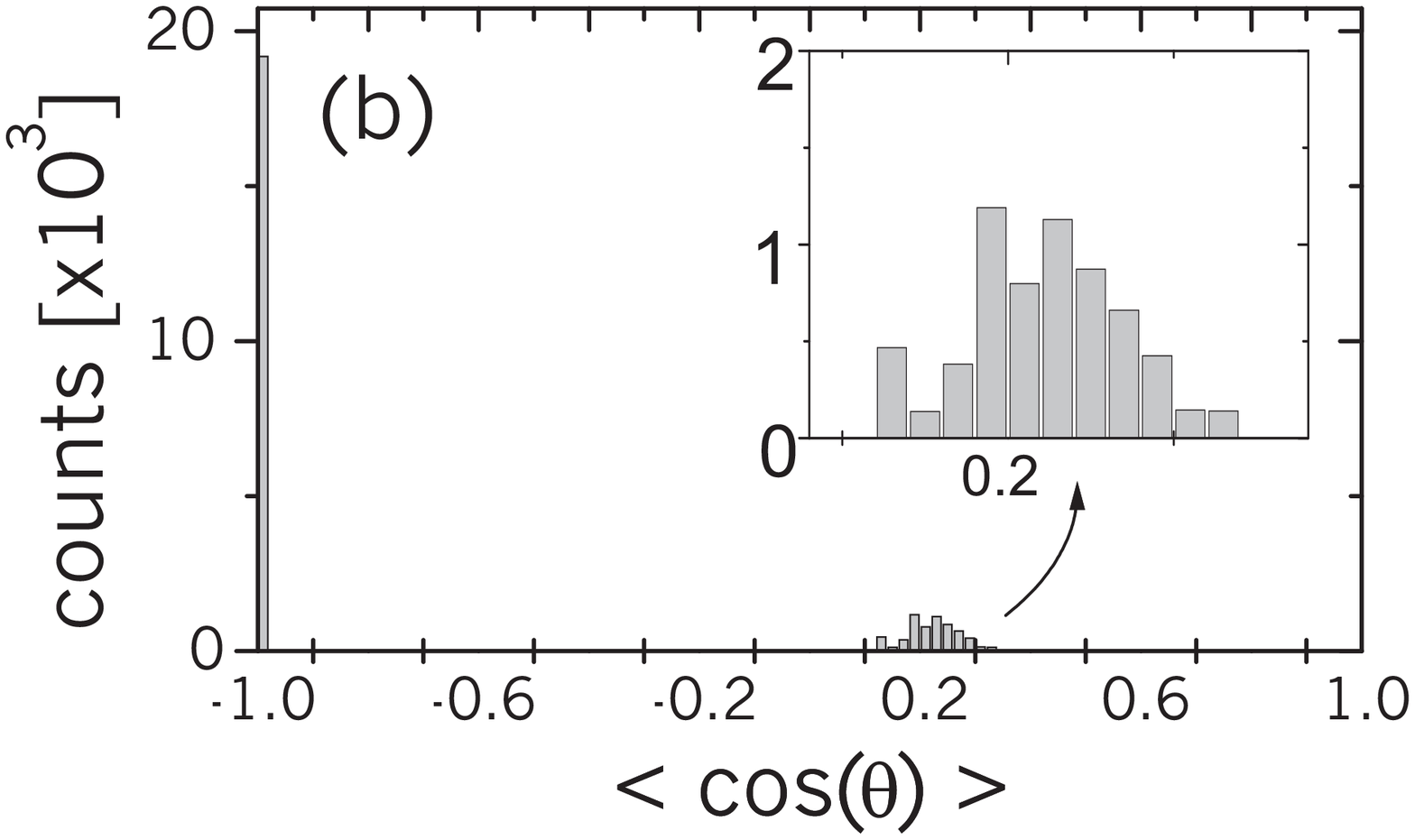,height=4.25cm,angle=0}}
\medskip
\centerline{\psfig{figure=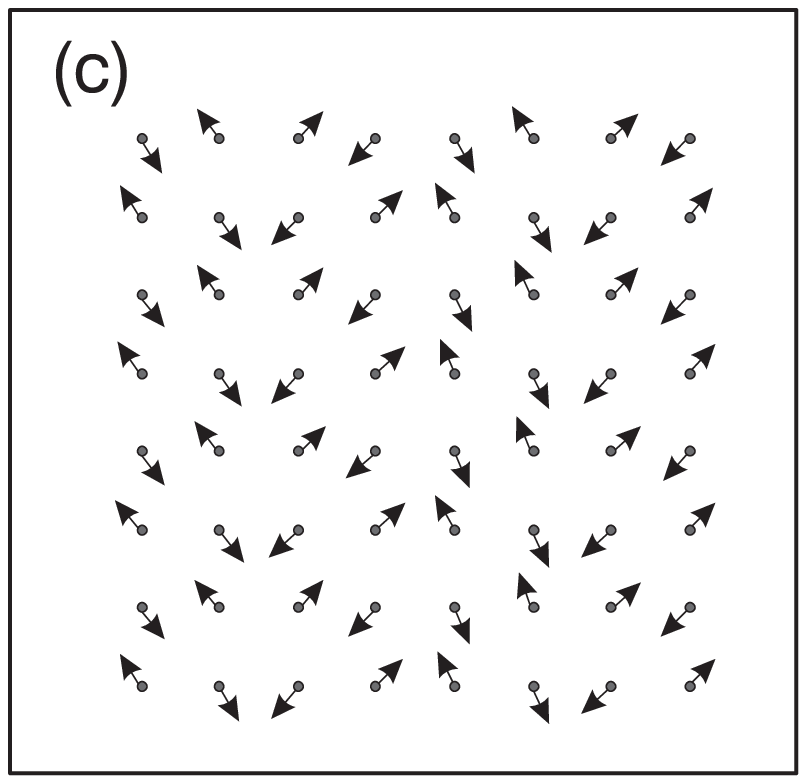,height=5cm,angle=0}}
\medskip
\caption{MC results for $n = 1/2$ at $K/t$ = 0.32, characterizing the 
$(\pi/2,\pi)$+$(\pi,\pi)$ phase on an 8$\times$8 lattice. (a) Structure 
factor. (b) Angle histogram. (c) Configuration snapshot. }
\end{figure}

Finally, for $n = 1/4$ the energy diagram (Fig.~5) in the region of 
small to intermediate $K/t$ is dominated by a single, and very robust, 
island phase. The extraordinary stability of the $(\pi/2,\pi/2)$ phase 
(Fig.~1(b)) at this filling is clear to see by diagonalizing 
the 4-site square cluster with hopping $t$. This exercise yields 
energy levels of $-2t, 0, 0, 2t$, the location of the gaps demonstrating 
immediately why the phase so favors 1/4-filling, but is so unfavorable 
at $n = 1/2$. Figs.~17(a)-(c) require little commentary, and we note 
only that the AF to FM angle ratio here is 1:1. As in Fig.~14, the 
charge-equivalence of all sites results in a homogeneous $n({\bf k})$ 
(Eq.~(\ref{ecdf})), but the delocalization of charge within the 
2$\times$2 squares would give a peak at $(\pi,\pi)$ in experiments 
measuring $n({\bf r})$.

\begin{figure}[hp]
\medskip
\centerline{\psfig{figure=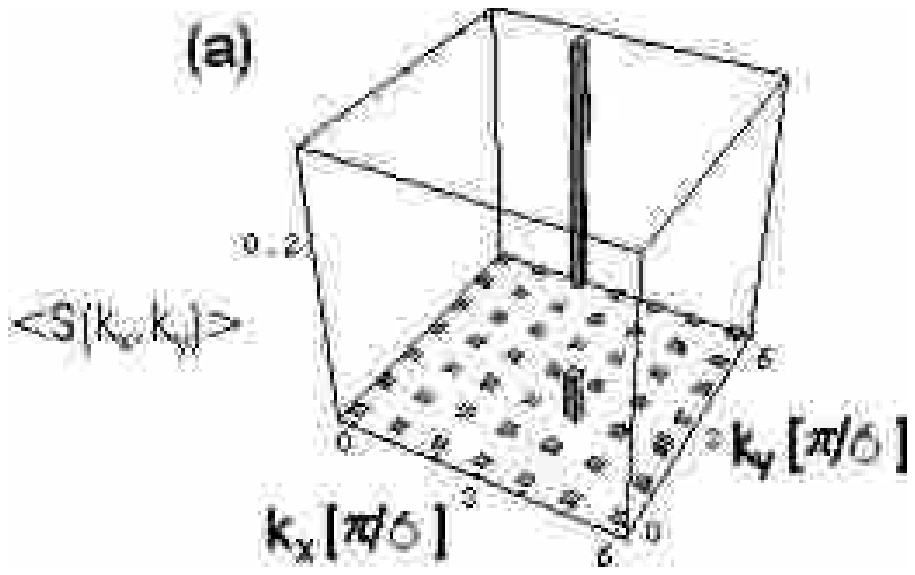,height=4.75cm,angle=0}}
\bigskip
\centerline{\psfig{figure=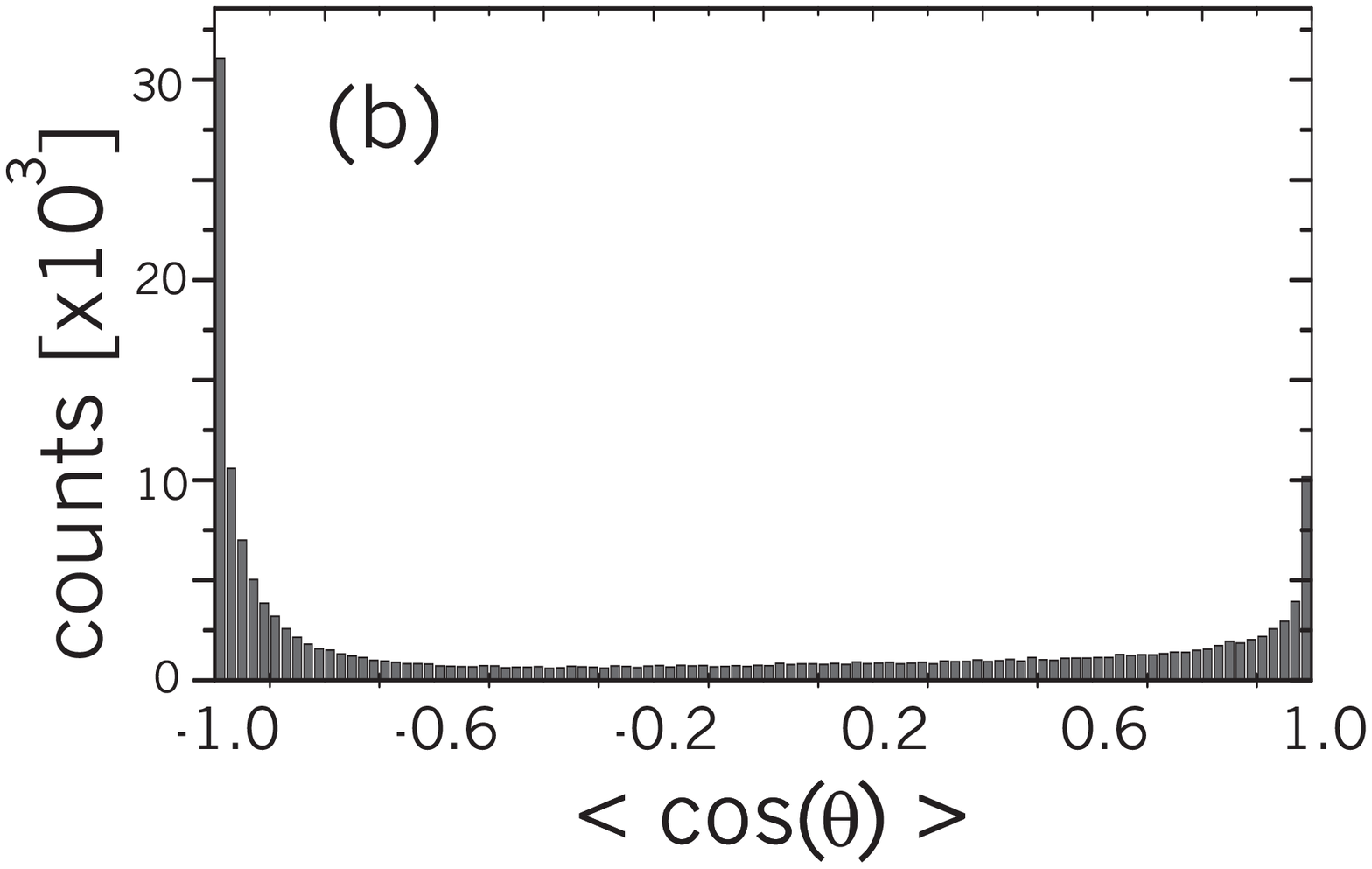,height=4.25cm,angle=0}}
\medskip
\centerline{\psfig{figure=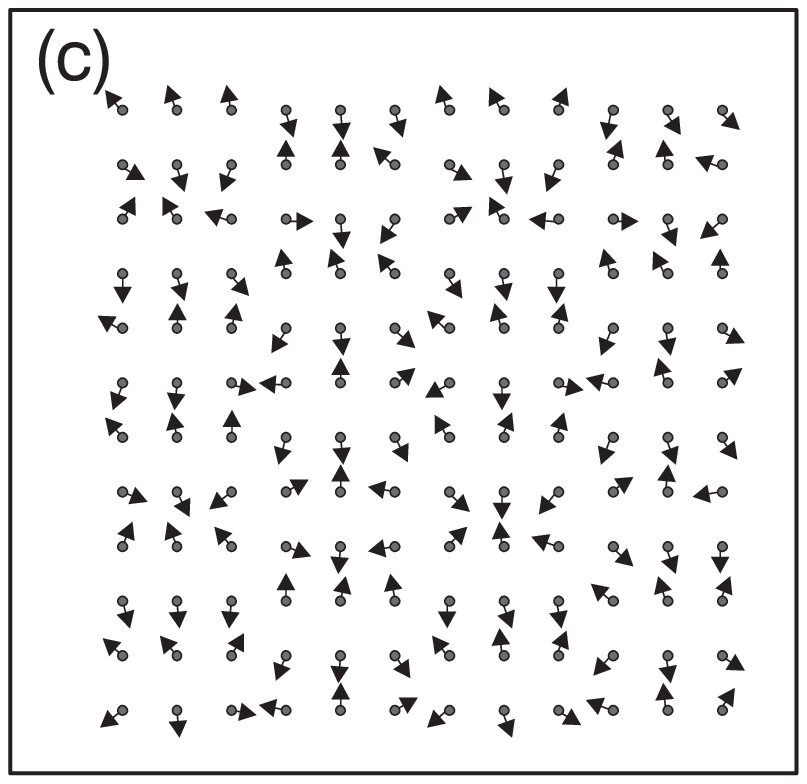,height=5cm,angle=0}}
\medskip
\caption{ MC phase for $n = 1/3$ at $K/t$ = 0.15, characterizing the 
$(\pi/3,\pi)$ phase on a 12$\times$12 system. (a) Structure factor. 
(b) Angle histogram. (c) Configuration snapshot. }
\end{figure}

Returning to the question of phase transitions, these may be considered 
as a function of $K/t$ or as a function of filling. In the former case 
the results are essentially those of Figs.~3-5. The only robust phases 
preceeding those in Figs.~13, 15, and 17 are FM phases, and 
at higher $K/t$ a short cascade of further states leads to the AF 
configuration. As described in Sec. III, the phases arising from MC 
simulations require a renormalization of their final energies to account 
for system size, and when this is performed the crossovers are fully 
consistent with the infinite-system results. In the experimentally more 
relevant case of fixed $K/t$ and variable filling, the results of Sec. 
IV imply that, for all but the smallest values of $K/t$, ``transitions'' 
take the form of a differential occupation of undoped and commensurately 
filled states, with the exception of the regime $0.2 < K/t < 0.28$ where 
they are replaced by a continuous evolution in the period of a 
large-unit-cell phase. 

\begin{figure}[hp]
\medskip
\centerline{\psfig{figure=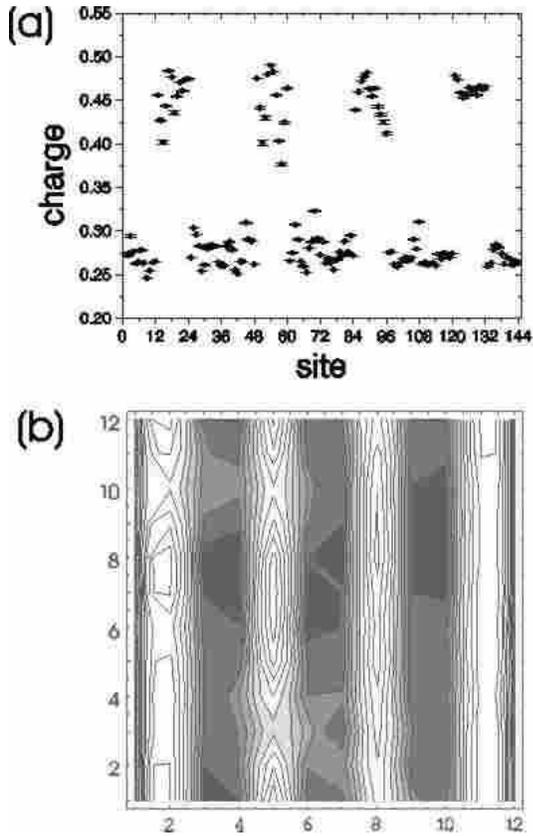,height=11cm,angle=0}}
\medskip
\caption{ Charge distribution function $n_i$ for $n = 1/3$ at 
$K/t$ = 0.15, illustrating charge order of $(\pi/3,\pi)$ phase on a 
12$\times$12 system. (a) Site charge densities: site numbers 1-12 label 
the first column from bottom to top (see (b), and Fig.~15(c)), 13-24 the 
second column from bottom to top, and so on. (b) Charge contour plot: 
high densities in white, low in grey. }
\end{figure}

Returning to the experiments presented in the introduction, our results 
justify certain, rather broad conclusions. Manganite systems which are 
structurally layered, or become layered as a result of orbital ordering 
in the cubic system, may indeed be susceptible to the island-phase 
phenomena, with resultant charge and spin order, discussed here. The 
fundamental ingredient for this is only the competition between $K$ and 
$t$ intrinsic to all materials in the class. However, we have emphasized 
throughout the crude nature of the model we consider, and close with a 
brief discussion of the possible extensions which may be required to 
reproduce more closely the physics of real materials. 

\begin{figure}[hp]
\medskip
\centerline{\psfig{figure=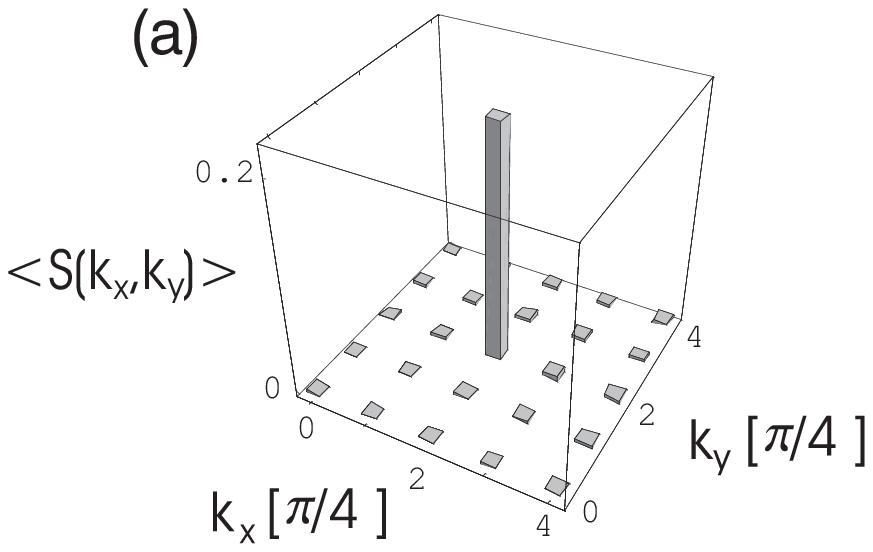,height=4.5cm,angle=0}}
\bigskip
\centerline{\psfig{figure=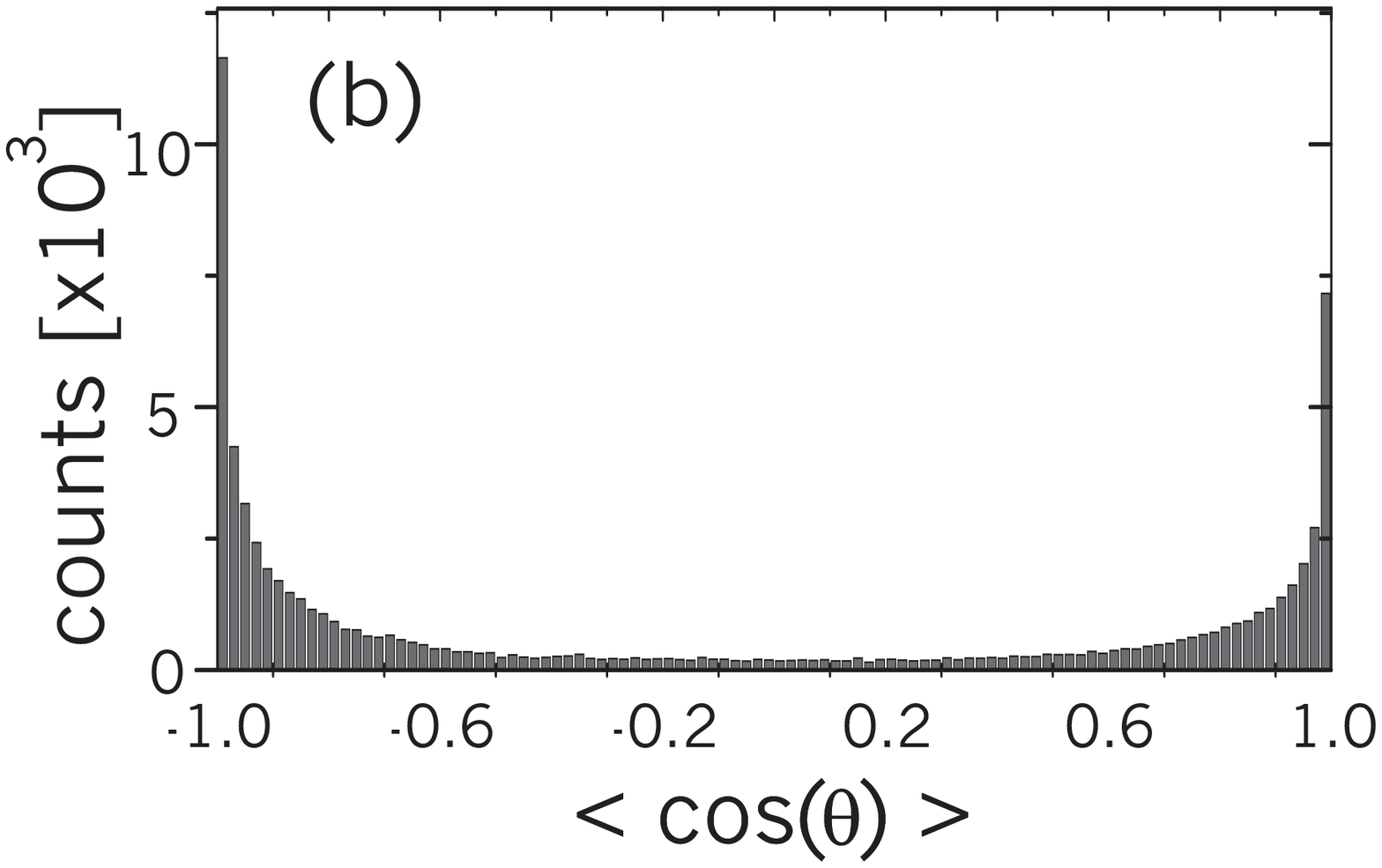,height=4.25cm,angle=0}}
\medskip
\centerline{\psfig{figure=ff15c.eps,height=5cm,angle=0}}
\medskip
\caption{MC phase for $n = 1/4$ at $K/t$ = 0.12, characterizing the 
$(\pi/2,\pi/2)$ phase on an 8$\times$8 lattice. (a) Structure factor.
(b) Angle histogram. (c) Configuration snapshot. }
\end{figure}

One of the fundamental features of manganite systems is the doubly 
degenerate nature of the $e_g$ orbital. This has been included by a 
number of authors, and has been argued\cite{rvk} to be essential in 
accounting for the CE-type (planar in 3d) charge order observed in 
La$_{1-x}$Sr$_x$MnO$_3$.\cite{rmcc} A further important ingredient in 
manganite systems is Jahn-Teller distortion of the local structural 
environment of each Mn ion,\cite{rcfg} which may act to lift the $e_g$ 
orbital degeneracy, and also to promote charge order. Both terms have 
been included in a classical MC study of the type performed here,\cite{rymd} 
albeit on very small systems. Island phases, in the orbital or spin degrees 
of freedom, were not among the already very rich variety of phases 
considered. When two $e_g$ orbitals are considered, on-site Coulomb 
interactions were found\cite{rhv} to lead to the formation of an 
upper Hubbard band, and to cause significant spectral weight shifts 
and broadening. Yet another term in many models of strongly-correlated 
electrons is a possible Coulomb repulsion between nearest-neighbor 
sites, conventionally denoted as $V$. This contribution acts to suppress 
phase separation, and to promote a charge ordering when $V$ competes with 
the hopping energy scale $t$, as noted in the 1d system.\cite{rghbaa} In 
higher dimensions, sufficiently strong $V$ may lead to anisotropic charge 
order if the hopping is anisotropic, and more generally for weak $V$ one 
expects a moving of phase boundaries to favor homogeneous states 
such as the stripes and islands considered here. Precisely this physics 
was found in Ref.~\onlinecite{rmyd}, where we note that the terminology 
``island'' phase is applied to mean a shrinking of the size of 
phase-separated regimes. We stress that the island phases and charge 
order in our study are intrinsic to the physics of the competing double 
exchange and superexchange, and that an additional $V$ term is not 
required for their appearance.

Finally, one of the major restrictions of the current approach is the 
limitation to small system sizes, which become smaller still on addition 
of the further terms discussed in the previous paragraph, and then still 
to largely classical considerations. The method of classical MC with 
diagonalization of the one-electron problem is in fact not particularly 
sophisticated, and we highlight here only two rather recent contributions 
which have the potential to reveal many more features on systems large 
enough to be considered thermodynamically representative. These are the 
variational mean-field \cite{rafglm1} and hybrid Monte Carlo\cite{rafglm2} 
techniques, both introduced for the double-exchange problem by the same 
group of authors, which allow extensions in the former case to 96$^3$ 
systems with appropriate approximations, and in the latter to 16$^3$ 
sites with rather fewer. A last important point is the question of 
corrections to the above results due to the effects of quantum fluctuations. 
In 1d, it was found\cite{rghbaa} that the boundaries between phases were 
moved to significantly larger values of $K/t$ than predicted classically. 
While the methods presented herein do little to allow an assessment of 
fluctuation effects, these should be significantly smaller in 2d, both 
directly because of the higher dimensionality, and because the 1d results 
were obtained with a localized ($t_{2g}$) spin $S$ = 1/2, whereas 
the classical limit may be no less representative of the physical 
situation ($S$ = 3/2). Thus our phase diagrams can be expected to be 
qualitatively quite accurate. One may also ask if quantum fluctuations 
would act to destroy the coherence of the large-unit-cell phases: 
because these phases are not spiral-ordered, and already possess the AF 
or FM local spin alignment favored by fluctuations, they may be assumed 
to be robust in this respect. 

\section{Summary}

In conclusion, we have considered the possibility of ``island'' 
phases and associated charge order in 2d systems, using as a model the 
augmented FKLM with strong Hund coupling. Indeed we find that stripe-like 
and island phases are stable at intermediate values of $K/t$ for each 
of the commensurate fillings $n =$ 1/2, 1/3 and 1/4. This result 
includes stability against global phase separation, even in the 
absence of additional Coulomb terms. Spiral magnetic order appears 
near the antiferromagnetic regimes at low filling or at large $K/t$. 
A variety of ``flux'' phases is possible, because the electron phase 
factor is non-trivial in all dimensions $d > 1$, but we find only 
one to be a stable ground state and this at $n = 1/2$. While the 
flux phase has a homogeneous charge distribution, the majority of 
the island phases show a charge modulation. Thus even the simple form 
(\ref{esh}) of the FKLM reproduces some of the most important experimental 
features of manganite charge and spin order. The critical values of 
$K/t$ for transitions between ordered phases, and between ordered and 
separated phases, may be identified rather accurately from classical 
considerations augmenting small-system studies. 

\section{Acknowledgements}

We are grateful to A. Aligia, C. Balseiro, C. Batista, D. Garcia, K. 
Held, and D. Poilblanc for helpful discussions. This work was supported 
by the Consejo Nacional de Investigaciones Cientificas y Tecnicas 
(CONICET) of Argentina, and by the Deutsche Forschungsgemeinschaft 
through SFB 484 (BN).

\end{document}